\documentclass{pasj01}

\begin{document}
\SetRunningHead{D.O. Gough }
{An Elementary Introduction to the JWKB Approximation}
\Received{2006/12/20}
\Accepted{2007/01/02}

\title{An Elementary Introduction to the JWKB Approximation}

\author{D. O. Gough}
\affil{Institute of Astronomy, Madingley Road, Cambridge, CB3 0HA, UK,}
\affil{Department of Applied Mathematic and Theoretical Physics,}
\affil{Wilberforce Road, Cambridge, CB3 0WA, UK, and}
\affil{Institut for Fysik og Astronomi, Aarhus Universitet, DK 8000 
\r{A}rhus C, Denmark}
\email{douglas@ast.cam.ac.uk}


\maketitle

\begin{abstract}
\noindent Asymptotic expansion of the second-order linear ordinary
differential equation $\Psi^{\prime\prime}+k^2f(z)\Psi=0$, in which
the real constant $k$ is large and $f={\rm O}(1)$, can be
carried out in the manner of Liouville and Green provided $f$ does not
vanish. If $f$ does vanish, however, at $z_0$ say, then
Liouville-Green expansions can be carried out either side of the
turning point $z = z_0$, but it is then necessary to ascertain how to
connect them. This was first accomplished by Jeffreys, by a comparison
of the differential equation with Airy's equation. Soon afterwards,
the situation was found to arise in quantum mechanics, and was
discussed by Brillouin, Wentzel and Kramers, after whom the method was initially
named. It arises throughout classical physics too, and is encountered
frequently when studying waves propagating in stars. This brief
introduction is aimed at clarifying the principles behind the method,
and is illustrated by considering the resonant acoustic-gravity
oscillations (normal modes) of a spherical star.
\end{abstract}

\section{Introduction}
It is common, and often apparently most straightforward, to approach
problems in macroscopic stellar physics by solving the governing
equations numerically. However, simple analytical techniques are often 
more revealing. For the latter it is usually necessary to idealize the
situation in hand, sometimes grossly so, to render
\,the equations tractable.  The outcome can reveal the phenomenon
under study in a very different light from that provided by specific
numerical examples. In particular, because in some respects analytical
results are more general, appreciation of their structure can guide
one more easily towards greater understanding. Even viewed merely 
as a diagnostic tool, that understanding, and sometimes just the 
analytical solutions themselves,\, have proved to be useful in the past 
simply for finding errors in numerical computations, both before and 
after publication. Numerical and analytical investigations are 
complementary, and each would be much poorer without the other.

This is the first of a short series of invited articles intended to
illucidate some of the analytical techniques that are used in macroscopic 
stellar physics. It discusses one of the most useful techniques for
studying the wave-like solutions of ordinary linear differential
equations of second order: namely, the so-called Liouville-Green
expansion combined with the method of Jeffreys for connecting
solutions across turning points, more commonly known as the WKB
approximation.  The method is presented and, I hope, made plausible, giving a
slight flavour of the background arguments without resorting to
mathematical proof, with the aim of guiding those not already
conversant with the method towards its proper use.

\section{The Liouville-Green expansion}
In modern times, the term WKB approximation has commonly been assigned
to what is more properly regarded as the leading term in the Carlini--Liouville--Green (LG) expansion of solutions of the second-order
ordinary linear differential eq\-uation
\begin{equation} \label{1.1}
\frac{{\rm d}^2\Psi}{{\rm d}z^2} + k^2 f(z)\Psi =0~,
\end{equation}
in which $k$ is real (positive) and large, and $f={\rm O}(1)$ is 
slowly varying, in the
sense that $k\gg|H^{-1}_f|\equiv |-{\rm d}\ln f/{\rm d}z |~$. Such equations
arise throughout classical physics in describing waves, and asymptotic
approximations to their solutions appear to have been considered first
by Carlini (1817), actually in a study of a problem in celestial
mechanics, and subsequently by Green (1837) and by Liouville (1837).

One of the simplest physical examples is that of small-amplitude adiabatic acoustic waves in an otherwise homobaric
fluid, whose linearized governing equations are
\begin{equation} \label{1.2} 
\rho_0 \frac{\partial^2\mbox{\boldmath$\xi$}}{\partial t^2}=- \nabla p^\prime~,
\end{equation}
\begin{equation} \label{1.3}   
p^\prime =\delta p =c^2\delta \rho = c^2\left(\rho^\prime
+ \mbox{\boldmath$\xi$}. \nabla \rho_0
\right) =
-\rho_0 c^2
\rm{div} \, \mbox{\boldmath$\xi$} ~,
\end{equation}
where $p=p_0+p^\prime$ is pressure, $\rho=\rho_0+\rho^\prime$ is
density and $\mbox{\boldmath{$\xi$}}$ is displacement, the zero denoting equilibrium
value, the prime, Eulerian perturbation, and $\delta$ denoting
Lagrangian perturbation. The square of the sound speed in the
unperturbed fluid is given by $c^2=\gamma_1 p_0/\rho_0$, where $\gamma_1$ is
the first adiabatic exponent. The unperturbed pressure  $p_0$ is
constant, and in this simple case $\rho_0$ and $c^2$ are considered to
vary with only a single Cartesian coordinate $z$, and not with
time, $t$.
The divergence of equation (\ref{1.2}) may be combined with equations 
(\ref{1.2}) and (\ref{1.3}) to yield the equation
\begin{equation} \label{1.4}  
\nabla^2 \delta p + H_{\rho}^{-1}\mbox{\boldmath$n$}. \nabla
\delta p - {1\over c^2} {\partial^2 \delta p \over \partial t^2} =0~,
\end{equation}
where $H_{\rho}^{-1}\mbox{\boldmath$n$}= - \nabla \rm{ln} \rho_0$ defines
the density scaleheight $H _{\rho}$, and $\mbox{\boldmath$n$}$ is a unit 
vector in the $z$ direction. In the special case in which the disturbance 
$\delta p$ varies only in the $z$ direction, equation (\ref{1.4}) reduces 
to the Klein-Gordon equation
\begin{equation} \label{KG} 
{\partial^2 \chi \over \partial t^2}+\omega^2_{\rm c} \chi = 
c^2 {\partial^2 \chi \over \partial z^2}~,
\end{equation}
where $\chi = \rho^{-{1\over 2}}_0 \delta p$ and 
\begin{equation} \label{1.6}
\omega^2_{\rm c}(z) = {c^2 \over 4 H^2_\rho} \left( 1-2 {{\rm{d}}H_\rho \over {\rm{d}}z}
\right)~.
\end{equation}

Because $\rho_0$ and $c^2$ are constant in time and are independent of
the co-ordinates perpendicular to $\mbox{\boldmath$n$}$, equation 
(\ref{1.4}) admits oscillatory solutions with frequency $\omega$ of the form 
$\delta p(\mbox{\boldmath$x$},t)=\Re
\left[\rho^{1\over 2}_0 \Psi(z)e^{i \left (\mbox{\boldmath$k$}_{\perp}
  . \mbox{\boldmath$x$}-\omega t\right )}\right ]$, 
where $\Re$ denotes real part and $\Psi$ satisfies
\begin{equation} \label{1.5}    
\frac{{\rm d}^2\Psi}{{\rm d}z^2}+\left(
  \frac{\omega^2-\omega^2_{\rm c}}{c^2}-k_{\perp}^2 \right )\Psi=0~,
\end{equation}
in which $k_{\perp}=\left|\mbox{\boldmath$k$}_{\perp}\right|$, where 
$\mbox{\boldmath$k$}_\perp$ is a constant wavenumber component
perpendicular to $\mbox{\boldmath$n$}$.  
Equation (\ref{1.5}) is essentially of the form of equation (\ref{1.1}), 
provided that $\omega$ is high enough.  

If $\rho_0$ and $c$ were actually constant, $\omega_{\rm c}$ would vanish
and equation (\ref{1.5}) would admit
solutions $\Psi=\Psi_0e^{\pm ik_{\parallel}z}$, where $\Psi_0$ is a constant
amplitude and 
$k_{\parallel}=\sqrt{\omega^2/c^2-k^2_{\perp}}$.
This represents a wave travelling with speed  
$v_\phi:=\omega/\left|\mbox{\boldmath$k$}\right|=c$ in the direction  
$\mbox{\boldmath$k$}= \left (
  \mbox{\boldmath$k$}_{\perp},\pm k_{\parallel} \right )$, provided that
$ k_{\parallel} $ is real,  because the 
phase $\tilde\phi=\mbox{\boldmath$k.x$}-\omega t$ is invariant in a frame
of reference moving in the direction of $\mbox{\boldmath$k$}$ with speed 
$v_\phi$, namely with 
$\mbox{\boldmath$x$}=\mbox{\boldmath$x$}_0 +v_\phi t
\hat{\mbox{\boldmath$k$}}$, where $\hat{\mbox{\boldmath$k$}}= 
\left|\mbox{\boldmath$k$}\right|^{-1}\mbox{\boldmath$k$}$.
The quantity $v_\phi$ is called the phase speed.

This solution motivates the LG expansion. Suppose first that $f>0$ in 
equation (\ref{1.1}).   If $c^2$ and $\omega_{\rm c}^2$ in equation 
(\ref{1.5}) vary 
\newline 
slowly with $z$ one can write
$ \left ( \omega^2 - \omega ^2 _{\rm c} \right ) / c^2- k^2_{\perp} = k^2
\kappa^2 (z)$, where $k$ is constant and
$\kappa = f^{1/2}$ is a function whose magnitude is of order unity and whose
scaleheight $H_\kappa$ is much greater in magnitude than $k^{-1}$. In
other words, the equilibrium, background, state varies only little
over the characteristic lengthscale $k^{-1}$ of variation of the wave. 
It is only under such circumstances that the concept of a wave is readily 
interpreted.

Thus, one poses a wave-like solution to equation (\ref{1.5}) by
setting
\begin{equation} \label{1.7}
\Psi=\Psi_0(z)e^{i\tilde\phi(z,t)}=\Psi_0(z) e^{i(k\psi(z)-\omega t)}~,
\end{equation}
regarding $k$ as a large parameter. It is usual then to expand $\psi$ 
in inverse powers of $k$;  here I shall afford myself the 
flexibility of expanding both $\psi$ and $\Psi_0$:
\begin{eqnarray} \label{1.8}
\psi&=& \psi_0 +k^{-1}\psi_1 +k^{-2}\psi_2 + \ldots~,\nonumber\\
\Psi_0&=& \Psi_{0,0}+k^{-1}\Psi_{0,1}+k^{-2}\Psi_{0,2}+ \ldots~.
\end{eqnarray}
Substituting expressions (\ref{1.7}) and (\ref{1.8}) into equation
(\ref{1.5}), with $k^2\kappa^2$ in place of 
$ \left ( \omega^2 - \omega ^2 _{\rm c} \right ) / c^2- k^2_{\perp}$, 
and equating to
zero terms of like order, yields the sequence of equations:
\begin{equation} \label{1.8a}
\psi_0^\prime = \pm \kappa~, 
\end{equation}
\begin{equation} \label{1.8b}
2i\psi^\prime_0 \Psi^\prime_{0,0} +(i\psi_0^{\prime\prime}
-2\psi_0^\prime\psi^\prime_1)\Psi_{0,0} =0~,
\end{equation}
\begin{eqnarray} \label{1.8c}
2i\psi^\prime_0\Psi^\prime_{0,1} &+&(i\psi_0^{\prime\prime}
-2\psi^\prime_0\psi^\prime_1) \Psi_{0,1} +\Psi_{0,0}^{\prime\prime}
+2i\psi^\prime_1 \Psi_{0,0}^\prime \nonumber\\
&+&(i\psi_1^{\prime\prime}-\psi^{\prime 2}_1
-2\psi_0^\prime \psi^\prime_2)\Psi_{0,0} =0~,
\end{eqnarray}
\begin{displaymath}
{\rm etc.,}
\end{displaymath}
where here and henceforth the prime denotes differentiation with
respect to the argument. These equations can be solved successively.

The first is equation (\ref{1.8a}), whose solution is
\begin{equation} \label{1.9a}
\psi_0=\pm \int \kappa {\rm d}z~.
\end{equation}
One now encounters in equation (\ref{1.8b}) a nonuniqueness, because
we have two new functions, $\Psi_{0,0}$ and $\psi_1$, with only one
equation to determine them. This is a result of the flexibility I
introduced by expanding the two functions $\Psi_0$ and $\psi$ in a
representation of only a single function $\Psi$, so there is
redundancy. One is therefore at liberty to choose either one of those
functions, or any relation between them, as one wishes. How one does
that can be regarded as a matter of convenience, and first I make the
common choice of setting $\psi_1=0$ and solving the resulting equation
for the leading-order amplitude, $\Psi_{0,0}$, yielding
\begin{equation} \label{1.9b}
\Psi_{0,0} \propto \psi_0^{\prime-1/2} \propto \kappa^{-1/2}~,
\end{equation}
the signs of proportionality indicating that one can multiply the
functions by any constant. One can continue, but evidently the
equations are beginning to become somewhat cumbersome; rather than
trying to write down a general procedure, it is preferable to tailor 
one's way through the complexity, using prudence to discern the way. 

Of course \, one could cut the cackle \, by setting $\psi_n =0$, 
\newline
$n>0$,
yielding
$2\psi^\prime
\Psi^\prime_{0,n}+\psi^{\prime\prime}\Psi_{0,n}- i\Psi^{\prime\prime}_{0,n-1}=
0,~\psi =\psi_0$,
 whence
 \begin{equation} \label{1.9c}
\Psi_{0,n}={\textstyle\frac{1}{2}}i\kappa^{-1/2}\int\kappa^{-1/2}~
\Psi^{\prime\prime}_{0,n-1}{\rm d}z~;
\end{equation}
this has the formal advantage of generating the entire sequence of
functions in one compact formula.   However, the sequence does not
always converge. Evidently, if one expands only $\Psi_0$ and if $\psi$ does not describe the phase
adequately, the (real part of the) exponential $e^{ik\psi}$ will
vanish in the wrong place, and although $\Psi$ might be forced to have
a zero in the right place in an attempt to rectify that, it cannot
remove a zero in $\Re (e^{ik\psi})$ that is in the wrong place without
itself being singular at that point. 
Therefore, the radius of convergence of the expansion is likely to be
smaller than it need be.  
Alternatively, one could expand only
$\psi$, as was common in the early days.  The procedure is 
algebraically  more complicated, but it tends to deliver a more robust 
approximation.  Alternate terms are real and 
imaginary, the latter accounting for the variation of the amplitude.  
In this respect, it is interesting to observe that if instead of
setting $\psi_1=0$ in equation (\ref{1.8b}) one sets $\Psi_{0,0}=1$, 
then $\psi_1^\prime=i\psi_0^{\prime\prime}/2\psi_0^\prime$; whence  
$\psi_1=\frac{i}{2}\ln\kappa$, and 
\begin{equation} \label{1.10}
\Psi \sim e^{\pm ik\int \kappa {\rm
d}z-\frac{1}{2}\ln\kappa}=\kappa^{-\frac{1}{2}} e^{\pm ik\int \kappa
{\rm d}z}~,
\end{equation}
which reproduces the solutions implied by equations (\ref{1.9a}) and
(\ref{1.9b}). Proceeding further, one obtains the second-order 
correction to the phase, which is given by equation (\ref{1.9d}) below; 
then the third-order term provides the second-order correction to the 
amplitude, which is identical to equation (\ref{1.9e}).

The algebraic technicalities are eased by expanding both $\psi$ 
and $\Psi_0$. 
One encounters nonuniqueness at each stage in the sequence of
equations (\ref{1.8a})--(\ref{1.8c}), and it is often expedient to
alternate between expansions in $\Psi_0$ and $\psi$, which is not surprising  
given that the terms in the phase expansion alternate between being real and 
imaginary when $\Psi_0$ is held fixed.  Thus, one can
set $\Psi_{0,1}=0$ in equation (\ref{1.8c}) and obtain
\begin{equation} \label{1.9d}  
\psi_2
=\pm {\textstyle\frac{1}{2}}\int\kappa^{-1/2}\left(\frac{{\rm
d}^2}{{\rm d}z^2}\kappa^{-1/2}\right){\rm d}z~, 
\end{equation}
and then set $\psi_3=0$ in the subsequent equation in the sequence to
yield
\begin{equation} \label{1.9e}
\Psi_{0,2}=-{\textstyle\frac{1}{4}}\kappa^{-2}\frac{{\rm d}^2}{{\rm d}z^2}\kappa^{-1/2}~,
\end{equation}
and so on. \,\, Equations (\ref{1.9d}) and (\ref{1.9e})\, are somewhat 
simpler to derive by this procedure than are their counterparts in the pure 
phase expansion.  The expressions for $\psi_2$ and $\Psi_{0,2}$ are 
essentially first and second derivatives of $\kappa^{-1/2}$ respectively; 
higher-order `corrections' to the solution contain
yet higher derivatives, which generically augur eventual divergence:
the expansion is asymptotic and must be terminated at some order.

If one needs to develop connexion formulae of higher-order than those
presented in \S5 it is prudent to arrange for $\Psi_{0,n}$ and
$\psi_{n+1}$ to be related in such a way as to ensure that the
Wronskian $\Psi_+ \Psi^{\prime}_{-} - \Psi_+^{\prime} \Psi_{-}$ of the
approximations $\Psi_+$ and $\Psi_-$ to linearly independent solutions 
of equation (\ref{1.5}) is
constant (Fr\"oman and Fr\"oman, 1996), as it is for the exact solutions, 
and as it is also for each of the asymptotic representations (\ref{1.10}) and 
(\ref{1.11}).

Terminating the LG expansion at equation (\ref{1.9b}) provides the
simplest, leading-order, representation. It results from solving just
the first pair of equations in the sequence beginning with equations
(\ref{1.8a})--(\ref{1.8c}). It is that formula which nowadays is often
called the WKB approximation.

In cases where $f<0$ in equation (\ref{1.1}), $\kappa^2 < 0$ in 
equation (\ref{1.5}),  solutions can be
obtained by exactly the same procedure as the wave-like solutions,
provided $|f(z)|$ remains of order unity so that the ordering
of the sequence of equations (\ref{1.8a})--(\ref{1.8c}) etc. is
preserved. The outcome in the so-called WKB approximation is again
given by equation (\ref{1.10}), which is better rewritten:
\begin{equation} \label{1.11} 
\Psi\sim |\kappa|^{-1/2}e^{\pm k\int|\kappa |{\rm d}z}~.
\end{equation}
The solutions therefore vary exponentially. 
Such `waves' are termed evanescent; evidently they do not propagate 
in the $z$ direction.

Near a point where $\kappa^2$ vanishes, the ordering of the sequence
of equations (\ref{1.8a})--(\ref{1.8c}) etc. is not preserved, and the 
approximation cannot be used.

\section{Normal form}
Equation (\ref{1.1}) is in what is called  normal form,
having no term in which $\Psi$ is singly differentiated. It might
appear at first sight that to add such a singly differentiated term 
would be of no serious
consequence, because the formal LG expansion could still be applied to
the more general equation. However, to do so is not prudent, for
unless one is very careful indeed, and perhaps even if one is very
careful, one risks ending up with a representation of the solution
whose domain of applicability is more restricted than it need
necessarily be. To illustrate the point let us consider again an
equation with constant coefficients:
\begin{equation} \label{2.1}
\frac{{\rm d}^2\Psi}{{\rm d}z^2}+2\eta \frac{{\rm d}\Psi}{{\rm
d}z}+k^2\Psi=0~,
\end{equation}
with $\eta={\rm O}(1)$. Its solutions are $\exp(-\eta z\pm
i\sqrt{k^2-\eta^2}z)$. Applying the LG expansion in the usual way, up
to ${\rm O}(k)$, yields the WKB equations:
\begin{equation} \label{2.2}
\psi_0^\prime=\pm k~,
\end{equation}
\begin{equation} \label{2.3}
\frac{\Psi^\prime_{0,0}}{\Psi_{0,0}}=-\eta~;
\end{equation}
whence
\begin{equation} \label{2.4}
\Psi\sim e^{-\eta z\pm ikz}~. 
\end{equation}
This approximate solution is valid only for a range $d$ of
$z$ satisfying $d\ll 2k/\eta^2$.

If, on the other hand, one writes $\Psi=e^{-\eta z}u(z)$, then 
\begin{equation} \label{2.5}
\frac{{\rm d}^2u}{{\rm d}z^2}+(k^2-\eta^2)u=0~,
\end{equation}
which is in normal form. Now the WKB-approximate solution is
actually exact. Of course, one could continue the direct LG expansion
of equation (\ref{2.1}) to higher order, which simply provides the
expansion of \,$\sqrt{k^2-\eta^2}z$\, in inverse 
powers of $k^2$, but it
takes the (formal) solution of two of the differential equations in the
analogue of the sequence (\ref{1.8a})--(\ref{1.8c}) etc. for each term,
which is a lot of work. Therefore it is undoubtedly at least expedient
first to cast equations with singly differentiated dependent variables
into normal form before proceeding with the expansion.

Of course, it could be said that the demonstration with an equation with
constant coefficients proves nothing about expansions of equations
with non-constant coefficients. 
\linebreak That is not wholly the case, however,
because provided the coefficients vary smoothly, so do the solutions,
and one can say that the casting into normal form is advantageous at
least for equations whose coefficients are close to being
constant. Furthermore, we have a great deal of experience with equations
of this kind, and we know that under a wide variety of circumstances
these asymptotic techniques based on small departures from constancy
work much better when the departures are not so small than perhaps one 
might feel they ought. This is not mathematical proof, but
pragmatism, based upon which I strongly recommend the taking of the
trouble to formulate the problem sensibly at the outset. To do so is
unlikely to cause a deterioration in the eventual outcome, and I 
baldly  assert that it is actually very likely
to provide substantial improvement.

Second-order linear ordinary differential equations can always be cast
into normal form, just as they can always be case into self-adjoint
form. If one has, for example, the equation
\begin{equation} \label{2.6}
\frac{{\rm d}^2y}{{\rm d}z^2}+2\eta\frac{{\rm d}y}{{\rm
d}z}+k^2\kappa^2 y=0~,
\end{equation}
where $k$ is constant, and $\eta$ and $\kappa$ are functions of $z$,
and if one wishes to preserve the independent variable $z$, \, then one
writes $y=u\Psi$, substitutes into the differential equation, and
simply chooses $u$ in such a way as to make the coefficient of ${\rm
d}\Psi/{\rm d}z$ vanish. The result is $u=\exp(-\int\eta {\rm d}z)$
and
\begin{equation} \label{2.7}
\frac{{\rm d}^2\Psi}{{\rm d}z^2}+\left(k^2\kappa^2-\eta^2-\frac{{\rm
d}\eta}{{\rm d}z}\right)\Psi=0~.
\end{equation}

It is also sometimes desirable to transform the independent variable
into something more natural, such as acoustic radius (namely, sound
travel time from the centre of the star), for example, if one were 
studying stellar acoustic waves. \ \ \  That
alone would destroy the normal form, if the equation had been in
that form in the first place. However, one can again start from
equation (\ref{2.6}), but first express it in terms of a new
independent variable $x$, and once again write $y=u\Psi$, this time
with $u=({\rm d}z/{\rm d}x)^{1/2}\exp(-\int\eta\frac{{\rm d}z}{{\rm
d}x}{\rm d}x)$, to obtain
\begin{equation} \label{2.8}         
\frac{{\rm d}^2\Psi}{{\rm d}x^2}+\left[k^2\kappa^2\left(\frac{{\rm
d}z}{{\rm d}x}\right)^2 -v(x)\right]\Psi=0~,
\end{equation}
where
\begin{equation} \label{2.9} 
v=w^2+\frac{{\rm d}w}{{\rm
d}x}~,~~~w(x)=\eta\frac{{\rm d}{z}}{{\rm d}x}-\frac{1}{2}\frac{{\rm d}}{{\rm d}x}
{\rm{ln}}\left(\frac{{\rm d}{z}}{{\rm d}x}\right). 
\end{equation}
In the case of waves propagating in the $z$ direction, it can be useful to 
replace $z$ by the mass variable $q=\int\rho~{\rm d}z$.  Waves in either a 
homobaric fluid or in a plane-parallel atmosphere 
stratified under constant gravity $g$ satisfy 
\begin{equation} \label{WE}
{\partial^2 \delta p \over \partial t^2}= 
\tilde c^2{\partial^2 \delta p \over \partial q^2}~,
\end{equation}
where $\tilde c =\rho c$, which establishes a transformation between the 
Klein-Gordon equation (\ref{KG}) and the wave equation (\ref{WE}). By applying 
the WKB approximation to the usual wave-equation analysis it is 
straightforward to demonstrate that an arbitrary infinitesimal disturbance
$\delta p$ propagates at a rate  $\tilde c$ along the $q$ coordinate in either 
direction without significant change of shape and with amplitude varying as
$\tilde c^{1/2}$, provided that the scale of 
variation of that disturbance is much less than the scaleheight of $\tilde c$. 
If the disturbance varies sinusoidally with time, the wave equation (\ref{WE}) 
reduces directly to the form (\ref{1.1}).

\section{Critical acoustic cutoff frequencies}
Before proceeding to a discussion of the JWKB approximation, which I
intend to illustrate with the problem of determining asymptotic
properties of acoustic-gravity waves, I pause for a moment to discuss
the so-called critical cutoff frequencies.\,\, They represent \, the frequency
\, beneath which 
\newline
a wave \, cannot \, propagate, \,\, although,\,\, when \, described in
\newline
these physical terms, it must be appreciated that they are not 
uniquely 
defined. That is simply because the concept of propagation itself is
not well defined. As I alluded at the outset, even the idea of a wave
is itself an asymptotic concept, and is not easily interpreted unless
the scale of variation of the background state is substantially
greater than the magnitude of the inverse wave number -- the very
conditions under which the approximations described here are designed
to be used. It has been a consequence of the resulting imprecision
that some apparently unsuspecting workers have been too careless in
writing down what should have been precise equations to describe a
physical situation, naturally under idealized, yet well defined,
conditions, and have so degraded their inferences unnecessarily.

The critical frequency that is perhaps the most familiar in physics is
the plasma frequency, $\omega_{\rm p}$. Indeed, its very existence is
the reason why plasma is so named (I refer here to an ionized gas, 
not to the liquid component of blood, although the basis for the latter's
appellation is essentially the same.)  The advantage of this example
over that of acoustic-gravity waves is that it can be considered for
an infinite uniform plasma (in the absence of an imposed magnetic
field), in which Langmuir waves of frequency $\omega$ satisfy the
equation
\begin{equation} \label{3.1}   
\frac{{\rm d}^2\Psi}{{\rm d}z^2}+\frac{\omega^2-\omega^2_{\rm
p}}{c^2}\Psi=0~,
\end{equation}
with $c^2$ and $\omega^2_{\rm p}=n_{{\rm e}}e^2/m_{\rm e}\epsilon_0$  
each being constant. 
In that case the critical cutoff frequency $\omega_{\rm p}$ 
is quite well defined in the physical terms I used above, as is the
concept of propagation: high-frequency waves, with $\omega
>\omega_{\rm p}$, propagate in the $z$ direction (which, of course, is
arbitrary) with wavenumber
$k=\sqrt{\omega^2-\omega^2_{\rm p}}/c$, whereas temporally periodic
disturbances with $\omega <\omega_{\rm p}$ are evanescent, having an
e-folding length $c/\sqrt{\omega^2_{\rm p}-\omega^2}$. 

I might mention, in passing, that Langmuir waves have been likened to the
Jeans waves in an essentially infinite uniform self-gravitating fluid,
a situation with\, which many\, astronomers are more familiar. \,These
waves are said to satisfy
\begin{equation} \label{3.2}  
\frac{{\rm d}^2\Psi}{{\rm d}z^2}+\frac{\omega^2-\omega^2_{\rm
J}}{c^2}\Psi=0~,
\end{equation}
(they actually do, but only approximately, for to avoid the so-called
Jeans swindle \,the background state must vary, slowly, in either space
or time, as Jeans himself recognized), where 
$\omega^2_{\rm J}=4\pi G\rho_0$. However, with respect
to scales much smaller than the Jeans length $c/\omega_{\rm J}$, the
gravitational term is negligible, and equation (\ref{3.2}) loses its 
cutoff.  It is only when
$\rho_0$ or $c^2$  varies with $z$ that a cutoff is reintroduced, 
but then the spatial variation tends to cloud the idea of propagation. 
The resulting imprecision is emphasized by comparing the Klein-Gordon equation 
(\ref{KG}), which contains an explicit cutoff frequency $\omega_{\rm c}$, 
with the precisely equivalent wave equation (\ref{WE}), which does not. 
The issue at stake, if one wishes to retain the physical
picture, is to define, in a spatially varying medium, where the wave
can propagate and where it cannot. I hope it is evident now that that
is unlikely to lead to a universally well defined end. But it can be
well defined under restricted circumstances, restricted in a sense
that I shall explain below. And provided that one confines oneself
consistently to the circumstances in which one has chosen to pose the
problem, a workable definition can emerge, and with it the chance
of a correct solution to the problem in hand.  Indeed, it is to the
task of determining how the solutions of equation (\ref{1.1}) in this
uncertain hinterland connect the well determined representations
(\ref{1.10}) and (\ref{1.11}) that the JWKB procedure is addressed.

Before proceeding, permit me to digress on the issue of the choice of
dependent variable.  It is not uncommon to work with the component
$\xi_{\parallel}:= \mbox{\boldmath$n$}. \mbox{\boldmath$\xi$} =:\xi_z$
of the displacement, or, equivalently, velocity, in the direction
$\mbox{\boldmath$n$}$ of variation of the background state, rather than
the (Lagrangian) pressure perturbation.  In the very simple case of the pure
acoustic wave considered in \S2, one first separates the parallel component 
$\xi_{\parallel}$ from the component of $\mbox{\boldmath$\xi$}$
perpendicular to $\mbox{\boldmath$n$}$, and then eliminates the
perpendicular component and $\delta p$ from equations (\ref{1.2}) and
(\ref{1.3}).  The procedure is straightforward, and yields
\begin{eqnarray} \label{3.2a}
\left(\nabla_\perp^2 - {1 \over c^2} {\partial^2 \over \partial
    t^2}\right)^2 \xi_{\parallel}
 +  H^{-1}_{c^2} \nabla_\perp^2  \mbox{\boldmath$n$}. \nabla
    \xi_{\parallel}\nonumber + \,\;\;\;\\ 
\left(  \nabla^2_{\perp}  -  {1 \over c^2}{\partial^2 \over \partial
    t^2}\right )   \left( \mbox{\boldmath$n$}. \nabla  - 
    H_{\gamma_1}^{-1}\right) \mbox{\boldmath$n$}. \nabla
  \mbox{\boldmath$\xi$}_{\parallel}= 0~, 
\end{eqnarray} 
in which $\nabla_\perp^2$ is the $\nabla^2$ operator in the plane
perpendicular to $\mbox{\boldmath$n$}; {\rm also} \ H{_{\gamma_1}} = \left (-
    \mbox{\boldmath$n$}. \nabla {\rm ln} \gamma_1 \right)^{-1}$ and $H_{c^2}
    = \left (- \mbox{\boldmath$n$}. \nabla {\rm ln} c^2 \right)^{-1}$ are the
      scaleheights of $\gamma_1$ and $c^2$, respectively.  
      This equation is rather
      more complicated than equation (\ref{1.4}).  In particular, it
      is of higher order, although after effecting 
the separation \newline $\xi_\parallel
      = \Re \left[ \Phi (z)e^{i
          \left(\mbox{\boldmath$k$}_\perp.\mbox{\boldmath$x$}-\omega t
          \right)}\right]$ the resulting ordinary differential
equation satisfied by $\Phi(z)$ is of only second order, 
as is equation (\ref{1.5}).
The corresponding equation pertaining to a background state that has
non-Cartesian symmetry, such as spherical symmetry, is yet more
complicated.  It provides justification for working with an intrinsic
scalar, such as $\delta p$, rather than the component of a vector.  I
remark also that in this simple case in which $p_0$ is constant, the
Eulerian and Lagrangian pressure fluctuations are numerically the same; I work
formally with $\delta p$ rather than $p^\prime$ because that 
generalizes more easily to the
situation in which the equilibrium state is stratified under gravity.

Let \,us \,consider \,now\, the \,adiabatic \,acoustic-gravity \,
waves in a spherical star. The governing equations in the Cowling 
approximation (an approximation obtained by neglecting the Eulerian 
perturbation to the gravitational potential) are
\begin{equation} \label{3.3}
\frac{{\rm d}\xi}{{\rm
d}r}+\left(\frac{2}{r}-\frac{L^2g}{\omega^2r^2}\right)\xi +
\left(1-\frac{L^2c^2}{\omega^2r^2}\right)\frac{\delta p}{\rho c^2}=0~,
\end{equation}
\begin{equation} \label{3.4}
\frac{{\rm d}\delta p}{{\rm d}r}+\frac{L^2g}{\omega^2r^2}\delta
p-\frac{g\rho F}{r}\xi=0
\end{equation}
(e.g. Gough, 1993) with respect to spherical polar coordinates 
$(r,\theta,\phi)$, where
$\xi$ is the vertical component of displacement from which has been
factored a spherical harmonic function $Y^m_l(\theta,\phi)$, $\delta
p$ continues to be the associated Lagrangian pressure perturbation,
but now with $Y^m_l$ factored out, $g$ is the local acceleration due
to gravity, with scaleheight $H_g$, and
\begin{equation} \label{3.5}
F=\frac{\omega^2r}{g}+2+\frac{r}{H_g}-\frac{L^2g}{\omega^2r}~,
\end{equation}
which I call the f-mode discriminant; $L^2=l(l+1)$, where $l$ is the
degree of the spherical harmonic that describes the angular variation
of the eigenfunctions. The other variables continue to retain the
meanings I assigned to them earlier. One can now eliminate $\xi$
from equations (\ref{3.3}) and (\ref{3.4})\, to yield \,a single
\,second-order \,dif\-fer\-en\-tial equa\-tion for\, $\delta p$, \,which, after
reduction to normal form, is approximated by
\begin{equation} \label{3.6}
\frac{{\rm d}^2\Psi}{{\rm d}r^2}+K^2\Psi=0~,
\end{equation} 
where $\Psi=(r^3/g\rho F)^{1/2}\delta p$ and 
\begin{equation} \label{3.7}
K^2=\frac{\omega^2-\omega^2_{\rm c}}{c^2}
-\frac{L^2}{r^2}\left(1-\frac{N^2}{\omega^2}\right)~,
\end{equation}
in which
\begin{equation} \label{3.8}
N^2=g\left(\frac{1}{H_\rho}-\frac{g}{c^2}\right)
\end{equation}
is the square of the buoyancy frequency.

Equations (\ref{3.6})--(\ref{3.8}) generalize equation (\ref{1.5}). They 
are not the exact equations to
result from equations (\ref{3.3})--(\ref{3.5}), but are what I call
the planar approximation to them, valid as $Kr
\rightarrow\infty$. They do not include the local effect of spherical
geometry, the only sign of sphericity that survives being the globally
geometrical representation $L/r$ of the horizontal
wavenumber. Including all the geometrical terms is straightforward
(Gough, 1993); they merely add a little complexity to the formulae
without changing those aspects of the mathematical structure of the
equation that concern us here, so I have omitted them for clarity. The
quantity $\omega_{\rm c}$, which is defined by equation (\ref{1.6}), 
is what is called the acoustic cutoff
frequency. It is not exactly a general cutoff frequency for $\Psi$ in
the sense that I described cutoff in connexion with equations (\ref{3.1})
and (\ref{3.2}), but is instead what that frequency would be for
spherically symmetric $(L=0)$ waves, uninfluenced by buoyancy.

Equation (\ref{3.6}) is similar to equation (\ref{1.1}), with
$k=\omega/\bar c$, $\bar c$ being a characteristic value of $c$ (such as the
value at the turning point, where $f=0$), and with $f$ depending on a
parameter $\alpha=\bar N/\omega$, where $\bar N$ is a characteristic value
of $N$. That view is appropriate for discussing acoustic
waves. Strictly speaking, equation (\ref{3.6}) is not precisely of the
form (\ref{1.1}), because $\alpha$ depends on $k$, but evidently $f$
becomes only very weakly dependent on $k$ as $k\rightarrow\infty$ and
the validity of the asymptotic arguments is unaffected. For gravity
waves one takes $k=L^2\bar N/\omega$ and $\alpha=\omega/\bar c$.

Forgive me for emphasizing at this stage what should be perfectly
obvious: there is no direct physical relation whatever between the
acoustic cutoff frequency $\omega_{\rm c}$ and the buoyancy
frequency $N$. The acoustic cutoff, which appears to have been
discussed first by Lamb (1909), arises when acoustic waves cannot
propagate vertically because the inverse wavenumber is \,comparable
with\, the density scaleheight; consequently there is inadequate
inertia on the low-density side of a compression to resist the
inevitable acceleration of matter, thereby annulling too much of the
pressure gradient to permit adequate subsequent compression of
the surroundings, essential for causing the perturbation to propagate
in a wave-like manner. The dynamics operates on the vertical component
of the motion, and is most effective for motion that is purely
vertical:  \ \ that motion has no \ horizontal variation. Buoyancy, on the
other hand, exists only when there is horizontal variation (cf.
Reye, 1872) and therefore $L$ must be nonzero, as is evinced
by equation (\ref{3.7}). One way of regarding it is to observe that
the force of gravity acting on a horizontally varying Eulerian density
perturbation is not in hydrostatic equilibrium, and the unbalanced
pressure gradients that are so engendered cause any typical fluid
element to be accelerated. That describes the
predominant dynamics of gravity waves. Confusion in the scientific
literature between the two totally different processes characterized
by $\omega_{\rm c}$ and $N$ appears to have arisen because, at least
in an isothermal atmosphere with constant $\gamma_1$, the formulae for
the two quantities can be made to look somewhat similar, and, if
$\gamma_1=5/3$, their values are almost the same, differing by only 4
per cent. That is no case for hiding the stark distinction between
them.

Permit me also to make another point which I hope by now is also quite 
obvious. There is a clear procedure for eliminating $\xi$ from
equations (\ref{3.3}) and (\ref{3.4}) --- one differentiates equation
(\ref{3.4}) to obtain an equation for ${\rm d}^2\delta p/{\rm d}r^2$
in terms of ${\rm d}\xi/{\rm d}r$ and $\xi$, and also ${\rm d}\delta
p/{\rm d}r$ and $\delta p$, then substitutes for ${\rm d}\xi/{\rm d}r$
using equation (\ref{3.3}), and then for $\xi$ using equation
(\ref{3.4}), leaving a second-order differential equation for $\delta
p$ -- and a well defined procedure for casting the resulting equation
into normal form, which I described in \S2, yielding a unique
dependent variable $\Psi$ (to within an inconsequential multiplicative
constant) and a unique equation (\ref{3.6}). Therefore the structure
of equations (\ref{3.6})--(\ref{3.8}) is well defined, and so
therefore is the acoustic cutoff frequency $\omega_{\rm c}$, and one
is not at liberty to change it. I hasten to add, however, that this
conclusion holds only within the restrictions I have imposed upon
myself: namely to use Lagrangian pressure perturbation (or, more
precisely, the appropriate multiple of it) as my dependent variable,
and radius $r$ as my independent variable.

The full critical cutoff frequencies $\omega_\pm$ associated with equation
(\ref{3.6}) for $\Psi$ are easily determined by factorizing $K^2$
(Deubner and Gough, 1984):
\begin{equation} \label{3.11}
c^2 K^2 =\omega^{-2}(\omega^2-\omega^2_+)(\omega^2-\omega^2_-)~,
\end{equation}
where
\begin{equation} \label{3.12}
\omega^2_\pm=\frac{1}{2}(S_{\rm L}^2 +\omega^2_{\rm c}) \pm
\left[\frac{1}{2}(S^2_{\rm L} +\omega^2_{\rm c})^2-N^2S^2_{\rm
L}\right]^{1/2}~,
\end{equation}
and where
\begin{equation} \label{3.13}
S_{\rm L}=\frac{Lc}{r}~,
\end{equation}
which is sometimes called the Lamb frequency. The situation is thus rather
more complicated than it is for pure acoustic waves, in which buoyancy
plays no part, and for pure gravity waves (in an incompressible fluid,
in which sound plays no part), which I have not discussed explicitly
here. Nevertheless, it is apparent from equation (\ref{3.11}) that
solutions resembling propagating waves of the form (\ref{1.7}) can be
found with $\psi$ real for waves with $\omega^2>\omega^2_+$ and for
waves with $\omega^2<\omega^2_-$, and indeed can be approximated by
the WKB solutions (\ref{1.10}) provided $K^2$ is large. If $\omega^2$
lies between $\omega^2_-$ and $\omega^2_+$, then $K^2<0$, the waves can
be regarded as being evanescent, and can be approximated by the
solutions (\ref{1.11}). What is most commonly encountered in practice
is a spatially varying yet temporally invariant background state, in
which the frequency $\omega$ of a wave is a conserved quantity, a
property which here I take for granted. There the wave can encounter
regions in which, for given $\omega$, $K^2>0$, and therefore
$\Psi^{\prime\prime}/\Psi<0$ -- the hallmark of a typical wave --
and regions in which $K^2<0$ and $\Psi^{\prime\prime}/\Psi>0$. They are
separated by well defined points at which $K^2=0$, and
$\Psi^{\prime\prime}=0$. It is therefore convenient to define the
former regions, quite precisely, as regions of propagation (often
abbreviated as propagating regions, even though the regions themselves
do not propagate), and the latter as regions of evanescence (or
evanescent regions). They are separated by the points at which
$\Psi^{\prime\prime}=0$, where the waves turn from one form to the
other; these points are called turning points. Generally, waves in the
propagating region cannot penetrate significantly beyond a turning
point. Yet the waves are very smooth there $(\Psi^{\prime\prime}=0)$,
so in reality unaccounted dissipation processes cannot be invoked to
destroy them, in contrast to the dynamics in the vicinity of a critical layer
(e.g. Booker and Bretherton, 1967), for example, where waves are absorbed. 
Therefore isolated turning points must be
points of total reflection. If there are two closely spaced turning
points enclosing an evanescent region, then, of course, barrier
penetration can occur, and reflection is not total.

Although the turning points of equation (\ref{3.6}) are defined
precisely, they do depend on the choice of both dependent and
independent variables. A different dependent variable, such as the
displacement, $\xi$, and, more pertinently, its associated counterpart
$\Xi$ that satisfies the normal form of the governing equation, is
not in phase with the Lagrangian pressure perturbation, and its points
of inflexion (together with the corresponding acoustic cutoff
frequency) must therefore be different. Indeed, so too do the local
vertical wavenumbers differ. But they are all well defined. Carrying
out the procedure corresponding to the derivation of equation
(\ref{3.6}) yields, again in the planar approximation,
\begin{equation} \label{3.14}
\frac{{\rm d}^2\Xi}{{\rm d}r^2}+K^2_\Xi\Xi=0~,
\end{equation}
where
\begin{equation} \label{3.15}
K^2_\Xi
=\frac{\omega^2-\omega^2_{\Xi{\rm c}}}{c^2}-
\frac{L^2}{r^2}\left(1-\frac{N^2_\Xi}{\omega^2}\right),
\end{equation}
in which 
\begin{equation} \label{3.16}
\omega^2_{\Xi{\rm c}}=\frac{c^2}{4H^2_{(\Xi)}}
\left(\!1\!+\!2 \frac{{\rm d}H_{(\Xi)}}{{\rm d}r}\right),
N^2_\Xi=g\!\left(\!\frac{1}{H_{(\Xi)}}-\frac{g}{c^2}\!\right),
\end{equation}
the mathematical structure of which is superficially similar to that
of equations (\ref{3.6})--(\ref{3.8}) and (\ref{1.6}). The scale  
$H_{(\Xi)}$ is defined according to
\begin{equation}
H^{-1}_{(\Xi)}=H^{-1}_\rho+
\left(1-\frac{L^2c^2}{\omega^2 r^2}\right)H^{-1}_{c^2};
\end{equation}
it depends on $\omega$,
rendering this formulation of the problem actually rather more complicated
than that in terms of $\delta p$. However, the formula for the 
corresponding acoustic
cutoff frequency $\omega_{\Xi{\rm c}}$, defined in the sense of being 
the cutoff frequency
for propagation of waves with $L=0$, as is the cutoff frequency
$\omega_{\rm c}$ defined by equation
(\ref{1.6}), is not dissimilar  (aside from a sign) to 
equation (\ref{1.6}), with $H_{(\Xi)}$
being instead the scaleheight of $\tilde c$. I have had to adorn the
vertical wavenumber $K$ (and the acoustic cutoff frequency $\omega_{\rm
c}$ and the buoyancy frequency $N$) with the subscript $\Xi$ to 
distinguish them from their
counterparts (\ref{3.7}) (and (\ref{1.6}) and (\ref{3.8})), to which, for 
consistency, I should attach the subscript $\Psi$.

A distortion of the independent variable also changes the locations of
the points of inflexion of $\Psi$ and $\Xi$, by an amount which is
defined by equations (\ref{2.8}) and (\ref{2.9}), although, of course,
they must always occur on the evanescent sides of the locations of the
nearest maxima to the evanescent regions. Transforming the independent
variable in equations (\ref{3.6}) and (\ref{3.14}) to acoustic radius
$\tau=\int c^{-1}{\rm d}r$, for example, yields
\begin{equation} \label{3.17}
\frac{{\rm d}^2 c^{\frac{1}{2}}\Psi}{{\rm d}\tau^2}+\left(c^2K^2_\Psi
-\omega^2_{\tau{\rm c}}\right) c^{\frac{1}{2}}\Psi=0
\end{equation}
and
\begin{equation} \label{3.18}
\frac{{\rm d}^2 c^{\frac{1}{2}}\Xi}{{\rm d}\tau^2}+(c^2K^2_\Xi 
-\omega^2_{\tau{\rm c}})c^{\frac{1}{2}}\Xi=0~.
\end{equation}
In each case the square of the appropriate acoustic cutoff frequency
is augmented by
\begin{equation} \label{3.19}
\omega^2_{\tau{\rm
c}}=\frac{1}{4T^2_c}\left(1-2\frac{{\rm d}T_c}{{\rm d}\tau}\right)~,
\end{equation}
where $T_c=(-{\rm d}\ln c/{\rm d}\tau)^{-1}$ is perhaps properly
called the sound-speed scaletime. Equation (\ref{3.18}) is analogous
to a similar equation presented by Christensen-Dalsgaard, Cooper and Gough
(1983) describing spherically symmetrical (radial) adiabatic
pulsations of a star with the perturbation to the gravitational
perturbation\  included \ (which for \ radial waves can be cast as a
second-order differential equation). If that equation is reduced to
the Cowling approximation, it agrees with equation (\ref{3.18}) with
$L=0$, as indeed it must.  These equations also reduce essentially to corresponding 
forms presented by Schmitz and Fleck (1998) in the case when $\gamma_1$ 
is assumed to be constant.

\section{The JWKB approximation}

Consider a wave, given approximately by equation (\ref{1.10}),
propagating to a turning point, beyond which it is evanescent. The
wave is therefore reflected, and travels back into the region of
propagation. \,An interesting \,and important question is: What is the
phase change, if any, on reflection? Viewed mathematically, one has
the two solutions (\ref{1.10}) well inside the region of propagation,
$z<z_0$ say, where $z_0$ is the location of the turning point, the
positive and negative signs in the exponent representing the incident
and the reflected wave respectively. In the evanescent region well
beyond the turning point, equation (\ref{1.11}) holds; here one must
choose the negative sign, because there is no disturbance far beyond
the point of reflection. The question can therefore be restated thus:
what combination of the two solutions (\ref{1.10}) in $z<z_0$ match
onto the decaying solution (\ref{1.11}) in $z>z_0$?  Readers not interested 
in the mathematical background to the answer to this question could skip 
the following subsection, save to accept the approximations 
(\ref{4.5})--(\ref{4.7}) to the two standard solutions Ai and Bi of Airy's 
equation (\ref{4.1}).

\subsection{Airy's equation}

The question posed at the end of the previous paragraph, in a somewhat 
different guise, had occupied the minds of mathematicians 
such as Stieltjes and Stokes in the mid 19th century.  In particular, the
young Stokes had wondered why it is that one cannot simply
analytically continue  representations 
(\ref{1.11})  -- actually, viewed explicitly as a representation of the 
Airy integral (\ref{4.2}) rather than a representation of the solutions of 
the differential equation (\ref{4.1}) -- 
 across  the turning point (to be more precise, around it, in the 
complex-$z$ plane, to avoid the singularity in the representation) into the 
representations (\ref{1.10}).  Although the WKB solutions (the
terminology I use here is evidently modern, for neither W, K nor B had yet 
been born) are invalid near the turning point, and therefore cannot be
properly connected on the real-$z$ axis, perhaps one might connect them
elsewhere in the complex plane along some contour $\cal C$ chosen to be far
enough from the turning point that $|kf|\gg 1$ everywhere on $\cal C$. That
it cannot be done to produce the correct
solution to the differential equation was subsequently named
the Stokes phenomenon. What Stokes wanted was to understand this
matter, and, of course, to obtain a connexion formula to link the two 
forms of solution. After several vain
attempts, he returned to the problem in 1857, and after three days of
concentrated effort the light (metaphorically) dawned at 3 o'clock in the 
morning.  Excitedly he wrote to his fianc\'ee on 19 March, telling her of 
his new realization, but prefacing it with a poignant acknowledgement of his
understanding that once they were married he would no longer be permitted 
to work to such hours\footnote{except occasionally}  (Larmor, 1907).

The matter is well illustrated by studying Airy's equation:
\begin{equation} \label{4.1}
\frac{{\rm d}^2y}{{\rm d}x^2}-xy=0~,
\end{equation}
which has a turning point at $x=0$, and whose solutions can be
expressed in terms of Bessel functions of order $\frac{1}{3}$. The
advantage of using such a simple equation, which, incidentally, is
central to the discussion of the more general equation
(\ref{1.1}) which follows, is that there are exact integral representations 
of its solutions which are analytic across the turning point. These are
\begin{equation} \label{4.2}
y= \frac{1}{2\pi i}\int_{\mathcal C} {\rm e}^{-xt+\frac{1}{3}t^3}{\rm d}t~,
\end{equation}
\begin{figure}
\includegraphics[width=80mm,height=80mm]{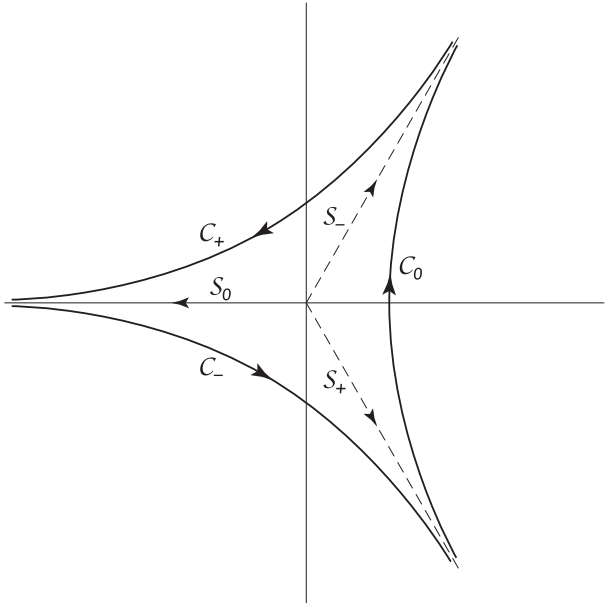}
\caption{The complex-$t$ plane divided into three regions by the straight 
lines $\mathcal{S}_-$, $\mathcal{S}_0$ and $\mathcal{S}_+$, each subtending
an angle of magnitude $2\pi/3$ with the others. The Airy integrals are
defined along the infinite contours $\mathcal{C}_-,~\mathcal{C}_0$ and
$\mathcal{C}_+$, each of which asymptote to $\mathcal{S}_-$,
$\mathcal{S}_0$ or $\mathcal{S}_+$ at infinity.}
\label{label1}
\end{figure}
the integral being along an appropriate contour $\mathcal{C}$ in the
complex-$t$ plane, which canonically is one of the curves
$\mathcal{C}_-,~\mathcal{C}_0$ or $\mathcal{C}_+$ depicted in
Fig. 1. There are various ways of deriving this result, but they are
not material to my presentation here. All that one need do is to
substitute the representation (\ref{4.2}) into equation (\ref{4.1})
and show that it fits. There are three natural solutions, $y_-$, $y_0$ and
$y_+$, represented by equation (\ref{4.2}) on the contours
$\mathcal{C}_-$, $\mathcal{C}_0$ and $\mathcal{C}_+$, respectively. Of
course they cannot be independent, because equation (\ref{4.1}) is of
only second order. Since the integrand is entire, it follows that the
integral along $\mathcal{C}_-\cup \mathcal{C}_0 \cup \mathcal{C}_+$ 
vanishes, and therefore
$y_-(x)+y_0(x)+y_+(x)=0$ for all (finite) $x$.

To span the solution space it is customary  to
adopt the two independent functions:
\begin{equation} \label{4.3}
{\rm Ai}(x)=y_0(x)~,
\end{equation}
evidently named after Airy (by Harold Jeffreys), and
\begin{equation} \label{4.4}
{\rm Bi}(x)=iy_-(x) - iy_+(x),
\end{equation}
a not unnatural consequent appellation. Both functions are real for
real $x$.
It is Ai$(x)$ which is of principal interest here, for that is the
solution that is relevant for our main purpose, for it decays,
exponentially, for large positive (real) $x$. It can be obtained in
terms of a single integral $I$ of a complex variable by deforming the
contour $\mathcal{C}_0$ to $(-\mathcal{S}_+)\cup\mathcal{S}_-$, on
which $t=e^{-i\pi/3}s$ or $t=e^{+ i\pi/3}s$ with $s$ real, whence
\begin{equation} \label{50}
y_0(x)=\frac{1}{2\pi
i}\left[e^{i\pi/3}I(e^{i\pi/3}x)-e^{-i\pi/3}I(e^{-i\pi/3}x)\right]~,
\end{equation}
which is evidently real when $x$ is real;
\begin{equation} \label{51}
I(z): \,=\int^\infty_0 e^{-zs-\frac{1}{3}s^3}{\rm d}s~.
\end{equation}
The function Bi may be expressed similarly.

Series expansions of the Airy integral (\ref{4.2}) 
 were developed by Stokes (1864, 1871); Jeffreys and
Jeffreys (1956) describe how to obtain asymptotic approximations for
large $|x|$ by Debye's method of steepest descents.  When $x>0$, the
saddle points in the complex-$t$ plane are at $t= \pm \sqrt{x}$, only
the one on the positive real axis being accessible to ${\cal C}_0$. In its
vicinity the line of steepest descents is parallel to the imaginary 
axis, and by deforming ${\cal C}_0$ to pass through $t= \sqrt{x}$ along 
that line one obtains immediately for ${\rm{Ai}}(x)$:
\begin{equation} \label{4.5}
{\rm Ai}(x)\sim\textstyle\frac{1}{2}\pi^{-1}x^{-\frac{1}{4}}\exp
\left(-\frac{2}{3}x^{\frac{3}{2}}\right) 
\left[1+{\rm O}\left(x^{-1}\right)\right]
\end{equation}
as $x\rightarrow +\infty$.  If $x<0$, the saddles are at $t= \pm
i\sqrt{-x}$, the lines of steepest descents being inclined at $\pm \pi
/ 4$ from the real axis, respectively. Neither is accessible to ${\cal C}_0$,
but we may evaluate $y_\pm$ instead and express $y_0$ as minus their sum: 
$y_\pm \sim \mp \left(2i \sqrt{\pi}\right)^{-1} (-x)^{-1/4}
\exp \left[\pm i \left(\frac{2}{3}(-x)^{3/2} +\frac{\pi}{4}\right)\right]$,\\
whence 
\begin{equation} \label{4.6}
{\rm Ai}(x)\!\sim\!
\pi^{-\frac{1}{2}}(-x)^{-{\textstyle\frac{1}{4}}}\!\sin\!
\left({\textstyle\frac{2}{3}} (-x)^{\frac{3}{2}}
+{\textstyle\frac{1}{4}}\pi\!\right)[1+{\rm O}(x^{-1})]
\end{equation}
as $x\rightarrow -\infty$.

The solution Bi$(x)$ grows exponentially as $x\rightarrow \infty$, as
of course does any other combination of $y_-$, $y_0$ and $y_+$ that
contains a nontrivial component of either $y_-$ or $y_+$. The
particular combination (\ref{4.4}) is chosen for defining Bi$(z)$
because it contains no exponentially small component in its asymptotic
expansion as $x\rightarrow +\infty$. 

The asymptotic analysis of ${\rm{Bi}}(x)$ for $x>0$ requires analysis in the 
vicinity of the saddle at 
$t= - \sqrt{x}$, through which the line of steepest descents is along the 
(negative) real axis.  Both $\cal{C}_+$ and $\cal{C}_-$ can be distorted to 
pass through it, which leads to a doubling of the amplitude factor:
\begin{equation} \label{4.6b}
{\rm Bi}(x)\sim \pi^{-\frac{1}{2}}x^{-\frac{1}{4}} \exp
\left({\textstyle\frac{2}{3}}x^{\textstyle\frac{3}{2}}\right) [1+{\rm
O}(x^{-1})]
\end{equation}
as $x\rightarrow +\infty$. For $x<0$ one appropriately combines the expressions 
for $y_-$ and $y_+$ obtained previously to yield
\begin{equation} \label{4.7}
{\rm Bi}(x)\!\sim\!
\pi^{-\frac{1}{2}}(-x)^{-\frac{1}{4}}\cos\left({\textstyle\frac{2}{3}}
(-x)^{\frac{3}{2}}\!+\!{\textstyle\frac{1}{4}}\pi\right)\![1+{\rm O}(x^{-1})]
\end{equation}
as $x\rightarrow -\infty$. Notice that the Wronskian of the 
asymptotic representations (\ref{4.5}) and (\ref{4.6b}), and (\ref{4.6}) and 
(\ref{4.7}) takes the same constant value, $\pi^{-1}$, either 
side of the turning point, as it should.

One can develop expansions to higher order, but I do not do so here.
I simply point out that for sufficiently large $x$ the relatively
small correction to expression (\ref{4.6b}) at any order exceeds even the
leading-order expression (\ref{4.5}).  Stokes' realization was that it is
because of that that one cannot analytically continue the
leading-order expressions around the turning point and expect them
still to represent the same solution of the differential equation (\ref{4.1}).

Note that expression (\ref{4.6}) provides the appropriate phase of the
oscillatory branch of the solution in this relatively simple case,
which was Stokes's goal. But it does not yet answer the question for
the more general equation (\ref{1.1}). However, one should perhaps pause 
for a moment to appreciate the power of the argument in the Airy case. 
Although the asymptotic expressions
(\ref{4.5})--(\ref{4.7}) are necessarily only approximations, they
are approximations to the exact representation (\ref{4.2}), and the
connexion between them is therefore robust. It isn't even necessary to
know what the functions look like for moderate or small values of $x$,
where the conditions motivating the asymptotic expansions at large
$|x|$ are not satisfied, although as a matter of natural curiosity one
might wish to know. Accurate numerical solutions satisfy that desire.
With such a connexion securely understood, Rayleigh (1912) applied the
analysis of the Airy integral to a study of the reflection of waves 
propagating through a medium (actually a stretched membrane, for which 
equation (\ref{1.5}) holds with $\omega_{\rm c}=0$) in which $c^{-2}$ 
varies linearly with $z$.

\subsection{Jeffreys' connexion}

The connexion between the asymptotic representations of the solutions
to the more general equation (\ref{1.1}) either side of a turning
point was first established by Jeffreys in his Cambridge Adams prize
essay in 1923 (see also Jeffreys, 1925). Subsequently, interest in the
issue arose in quantum mechanics with respect to solutions of the 
time-independent Schr\"odinger equation with a Coulomb potential, 
in studying, in particular, the
asymptotic energy levels of the hydrogen atom. 
\, Brillouin (1926) related Schr\"odinger's equation to 
Hamilton-Jacobi theory of classical mechanics, and Wentzel (1926)
discussed the leading-order LG expansion of Schr\"odinger's equation;
in response, Kramers (1926), who as an impoverished young man lived as a
guest of the Jeffreys in their house in order to make ends meet during
a stay in Cambridge, added 
the connexion formula that Harold Jeffreys had established. Mathematically
their analysis was not new, but, being
applied to a branch of physics that was more fashionable than the
classical wave propagation that had interested Rayleigh and Jeffreys,   
it attracted the
attention of more physicists (e.g. Young and Uhlenbeck, 1930; Kemble, 1935), 
and also mathematicians such as Langer (1934, 1937, 1949), and appellations
combining the initials of the surnames of the three quantum
physicists, in various orders, were given to the method, eventually
converging on the order WKB. Yet later, when Jeffreys' pioneering work
was more widely recognized, the initial J was added, either in front
or, more commonly, behind. I adopt the former, partly because Jeffreys
has precedence,   and partly because his analysis was designed to solve
a wider class of problems than \, merely\, those\, associated \, with \, the \, 
time-independent 
\newline Schr\"odinger equation. It is 
interesting to contemplate, however, that had those who originally
named this omnipresent and forgiving approximation been more
aware of its true history, they might have dropped the initials W, K
and B, and instead called it the SJ approximation\footnote{after
Stokes and Jeffreys}.

There are now several ways of justifying the approximation
(e.g. Heading, 1962). Jeffreys (1925) noted that if $f(z)$ has a
simple zero at $z=z_0$, namely $f(z_0)=0$ and $f^\prime(z_0)\not= 0$,
then $f(z)\simeq f^\prime(z_0)(z-z_0)$ near $z=z_0$; and if
$f^\prime(z_0)>0$, the substitution
$x=-k^{\frac{2}{3}}[f^\prime(z_0)]^{\frac{1}{3}}(z-z_0)$ transforms
equation (\ref{1.1}) approximately into Airy's equation (\ref{4.1}),
which enables approximate solutions of the full equation to be likened
to the exact solutions of the comparable approximate equation, and
hence to a connexion between representations of the solutions either
side of the turning point (via the integral representation
(\ref{4.2})). He pointed out that the outcome is a limiting form of
the true solution if $k$ is arbitrarily large. Thus the phase of that
oscillatory solution in $z<z_0$ that connects to the solution that
decays as $z\rightarrow +\infty$ is determined from equation
(\ref{4.6}).

Nowadays it is customary to use an approximation that is also valid
far from the turning point. This is achieved by applying what has
become known as the Liouville-Green transformation to equation
(\ref{1.1}), namely
\begin{equation} \label{4.8}
x=-k^{\frac{2}{3}}{\rm sgn}(f)\left|{\textstyle\frac{3}{2}}\int^z_{z_0}
f^{\frac{1}{2}}{\rm d}z\right|^{\frac{2}{3}}~,
\end{equation}
\begin{equation} \label{4.9}
\Psi=\sigma\Phi := (-xf^{-1})^{\frac{1}{4}}\Phi~,
\end{equation}
which leaves the equation in normal form. It may be written as
\begin{equation} \label{4.10}
\frac{{\rm d}^2\Phi}{{\rm d}x^2}-x\Phi=k^{-2}h(x)\Phi~,
\end{equation}
where
\begin{equation} \label{4.11}
h\{x[\sigma(z)]\}=-\sigma^3\frac{{\rm d}^2\sigma}{{\rm d}z^2}~.
\end{equation}
First it should be noted again that near the turning point $f\simeq
f^\prime_0(z-z_0)~,~ {\rm where}~f^\prime_0=f^\prime(z_0)$, and that
therefore $x\sim-{\rm
sgn}(f^\prime_0)k^{\frac{2}{3}}|f^\prime_0|^{\frac{1}{3}}(z-z_0)$,
which is Jeffreys' transformation: the term on the right-hand side of
equation (\ref{4.10}) arises from small higher-order terms in the
Taylor expansion of $f$ about $z_0$, and can be neglected. Far from
the turning point $f$ is no longer small, but, by definition, is of
order unity. One can therefore estimate the ratio of the right-hand
side of equation (\ref{4.10}) to the second term on the left-hand side
to be ${\rm O}(k^{-2}\sigma^4/xH^2_f)\!=\!{\rm O}(k^{-2}H^{-2}_f
f^{-1})={\rm O}(k^{-2}H^{-2}_f)={\rm o}(1)$ as
$k\rightarrow\infty$. $\!$There\-fore, provided that $k$ is
sufficiently large, the right-hand side can be ignored throughout,
reducing equation (\ref{4.10}) to Airy's equation. What I have
described here is not a proof, but it suggests that the
solution, $\Phi={\rm Ai}(x)$, to what is now called the comparison
(Airy) equation provides a valid approximation to the required
solution of equation (\ref{1.1}). And indeed Olver (e.g. 1974, 1978)
has shown that for sufficiently well behaved functions $f$ \,the
Airy-function approx\-imation converges uniformly for all $x$ 
to the solution of
equation (\ref{4.10}) as $k\rightarrow\infty$.

It should be recalled that the JWKB procedure described above applies to 
simple turning points, for which $f^\prime(z_0)\not= 0$.  If  
$f^\prime(z_0)=0$ but $f^{\prime\prime}(z_0)\not= 0$, then one can carry 
out a similar analysis using Weber's equation in the form 
${\rm d}^2 y/{\rm d}x^2$
\newline
 $+x^2 y=0$
as the comparison, and one may generalize further to yet higher-order
turning points. Olver has proved uniform convergence in such cases
too.

The uniformly valid JWKB approximation to the solution to equation
(\ref{1.1}) having a single simple turning point at $z=z_0$ such that
$f>0$ for $z>z_0$ can therefore be written in terms of ${\rm Ai}(-x)$,
where $x$ is defined by equation (\ref{4.8}). Far from the turning
point the appropriate asymptotic approximation (\ref{4.5}) or
(\ref{4.6}) to the Airy function is applicable. It yields,
after setting $f=\kappa^2$, as in \S1, and relating $\Phi$ to $\Psi$
by equation (\ref{4.9}),
\begin{equation} \label{4.12}
\Psi\sim {\textstyle\frac{1}{2}} A|\kappa
|^{-\frac{1}{2}}\exp(-\int^{z_0}_z|\kappa|{\rm d}z)~~{\rm for}~z\ll z_0,
\end{equation}
and
\begin{equation} \label{4.13}
\Psi \sim A\kappa^{-\frac{1}{2}}\sin \left(\int^z_{z_0}\kappa
{\rm d}z+\frac{\pi}{4}\right) ~~{\rm for}~z\gg z_0,
\end{equation}
where $A$ is a constant amplitude. These are equivalent to the
Liouville-Green expressions (\ref{1.11}) and (\ref{1.10}); but they
are more precise, because they define which combination of the
solutions (\ref{1.10}) corresponds to the evanescent solution in
$z<z_0$. It goes without saying that there is a similar pair of 
JWKB-approximate solutions that corresponds to exponential growth 
in the so-called
evanescent zone. Such solutions would need to be combined with
(\ref{4.12}) and (\ref{4.13}) if one were solving, for example, a
barrier-penetration problem, either classical or quantum.

\section{Equations with two well separated turning points: eigenvalue
problems}

Few sharp boundaries are encountered in astrophysics. Normally, in
wave problems, when waves are confined within some region
$\mathcal{P}$ it is because they encounter other, bounding, regions
$\mathcal{E}$ into which they cannot propagate, but the `interface' is
of finite thickness and is smooth. (I have in mind a 3-dimensional
space with propagating and evanescent regions $\mathcal{P}$ and
$\mathcal{E}$ separated by 2-dimensional surfaces.) That is not to say
that no wave at all can propagate in the bounding regions $\mathcal{E}$. Usually the
bounding regions can support waves of the same type as those under
consideration, but with rather different frequency or wave number. The
`boundary' $\mathcal{B}$ between those regions can usually be regarded
as the location of a turning point (or, more generally, as the locus
of a turning point), in the sense that I have used it in this article,
and is as well defined as are the critical cutoff frequencies.

I confine my discussion to situations in which the background state is
independent of time, so that frequency $\omega$ is well defined, and
is conserved, as is wave energy in a dissipationless system. (An LG
expansion in time, and even a JWKB approximation, can be made to study
waves in an appropriately slowly changing environment, but I do not
address that here.) Consider there to be a set of locally Cartesian
coordinates $(\xi,\eta,\zeta)$ established in the vicinity of the
boundary, with $\zeta$ perpendicular to the boundary. (The co-ordinate 
$\xi$ here is 
not to be confused with the vertical component of the displacement of 
equation (\ref{3.3}).)  Generically, the background (equilibrium) state 
varies more rapidly with $\zeta$ than it does with $\xi$ and $\eta$
(there are exceptions). On a length scale smaller than the scale of
variation on $\mathcal{B}$ of the background state, the differential
equation describing the waves reduces approximately to an ordinary
(linear) differential equation  with
respect to $\zeta$ (often of second order, analogous to
equation (\ref{1.1}), the only case I consider explicitly here), 
with a turning point `on' $\mathcal{B}$. Why does
that turning point arise?  In other words, why is wave propagation
possible in $\mathcal{P}$ but not in $\mathcal{E}$?

One reason might be that no wave of frequency $\omega$ can propagate
in $\mathcal{E}$, whatever its orientation. Then a wave incident on
$\mathcal{B}$ has its direction reversed -- equation (\ref{1.1})
permits no other possibility -- undergoing what is called (true)
reflection. Alternatively, $\mathcal{E}$ simply doesn't permit
continued propagation of a wave with a particular angle of incidence:
if the background state is independent of some particular coordinate
$\xi$ (or $\eta$, it cannot be $\zeta$), then the component
$k_\xi$ of the local wavenumber is conserved across the boundary, and
if $f$ in equation (\ref{1.1}) is then rendered negative in
$\mathcal{E}$ for that value of $k_\xi$ (at the frequency $\omega$ of
the wave), propagation is impossible; indeed, because the variation of
the background state is gentler in $\mathcal{B}$ than perpendicular to
it, that describes approximately the situation with respect to the
entire component $k_\perp$ of the wavenumber in $\mathcal{B}$ (i.e. 
perpendicular to the normal). The
process is often called total internal reflection, and, if $f$ is 
continuous, the case first considered by Rayleigh (1912), can be thought
of as the result of continuous refraction in the vicinity of
$\mathcal{B}$.   Note that because the exponents of the two LG solutions 
(\ref{1.10}) have the same magnitude at any point, the
angle of reflection is equal to the angle of incidence. Note also
that, because $f$ varies when either $\omega$ or $k_\perp$ vary (at
given $\zeta$), the location of $\mathcal{B}$ varies with not only
frequency but also with the angle of incidence of the wave; and, for
some values of $\omega$ and $k_\perp$, $\cal B$ may not exist.

A region $\mathcal{P}$ that is enclosed by $\mathcal{E}$ is called a
cavity. Any wave confined within it, in the absence of dissipation,
will, under most situations, eventually pass arbitrarily close to any
point it had passed formerly, and interfere with itself. For certain
frequencies, that interference is constructive everywhere in  
$\mathcal{P}$, and resonance
occurs. The resulting motion is called a mode of oscillation: the
contents of the entire region $\mathcal P$ oscillate in unison with frequency
$\omega$ (it being quiescent elsewhere). For sufficiently high $k$ the
LG expansion can be used along the phase trajectories of the waves,
using JWKB theory in the vicinity of the boundary of $\mathcal{P}$, 
to determine the
values of $\omega$ that permit resonance. These values are called the
eigenfrequencies of the cavity. Such an analysis can be geometrically
complicated enough to detract from the main point of this discussion,
so I postpone discussion of the general problem to another
article. Here I simply consider a problem in just one dimension, in which a
wave after reflection is bound to pass through all the points it had
passed through previously. Then the governing differential equation is
of the type (\ref{1.1}), with $k$ and $f$ depending in some way on
$\omega$.

To be more specific, \ let us consider a situation in which $f(z)=\kappa^2>0$
for $ z_1<z<z_2$, and $f<0$ elsewhere. Asymptotic solutions
$\Psi_2(z)$ in the vicinity of $z_2$, and far from $z_1$, may be
written as $\sigma(x)\Phi_2(x)$, where $x$ is defined by equation
(\ref{4.8}) with $z_0$ replaced by $z_2$, $\sigma$ is defined by equation
(\ref{4.9}), and $\Phi(x)={\rm Ai}(x)$. In the vicinity of $z_1$ one
may do likewise to define $\Psi_1(z)=\sigma(x)\Phi_1(x)$, but with
$x$ defined with the opposite sign, and now with $z_0$ replaced by
$z_1$, to ensure that the interval in which Ai is oscillatory
satisfies $z>z_1$. For simplicity I shall assume that $z_1$ and $z_2$
are sufficiently far apart that there is a common region of validity
in $(z_1,z_2)$ of the high-$(-x)$ expansion (\ref{4.5}) of $\Phi_1$
and $\Phi_2$. This must be possible for sufficiently large $k$. Then
$\Psi_1$ and $\Psi_2$ can be matched. According to equation
(\ref{4.13}),
\begin{eqnarray} \label{5.1}
\Psi_1\sim A_1\kappa^{-\frac{1}{2}}\sin\left(k\int^z_{z_1}
\kappa {\rm d}z +{\textstyle\frac{\pi}{4}}\right)~{\rm for}~z&\gg& z_1 \nonumber\\
({\rm and}~z &\ll& z_2)
\end{eqnarray}
and
\begin{eqnarray} \label{5.2}
\Psi_2\sim  A_2\kappa^{-\frac{1}{2}}\sin\left(k
\int^{z_2}_{z}\kappa {\rm d}z  +{\textstyle\frac{\pi}{4}}\right)~{\rm
for}~z&\ll& z_2 \nonumber \\ 
{\rm (and}~z&\gg& z_1)~.
\end{eqnarray}
The \,representations are identical throughout the common interval of 
validity if $A_2=\pm A_1$ and
\begin{equation} \label{5.3}
{\sin\atop\cos}\left({\textstyle\frac{1}{2}}k\int^{z_2}_{z_1}\kappa{\rm
d}z+{\textstyle\frac{\pi}{4}}\right)=0~,
\end{equation}
which is possible only if 
\begin{equation} \label{5.5}
k\int^{z_2}_{z_1}\kappa {\rm d}z=\left(n-{\textstyle\frac{1}{2}}\right)\pi~,
\end{equation}
where $n$ is an integer, and, in order to achieve a possible matching with 
the two evanescent solutions either side of the oscillatory
region, at least one half-wavelength is required: \,therefore
$n=1,2,3,\ldots~$. This is the desired eigenvalue equation.

I promised to illustrate the approximation with acoustic-gravity modes
of a star, for which, in the case of a spherical star,  
$k\kappa$ is $K$ given by equation
(\ref{3.7}). I do so with some hesitation, however, because stars are
not quite as simple as the situation I have just discussed. The reason
is that if $\omega$ is high, which is one way of making $K$ large,
then the upper turning point $r_2$, which occurs roughly where
$\omega^2_{\rm c}(r)=\omega^2$, is close to the surface, towards which
$\omega^2_{\rm c}$, and perhaps also $N^2$ (depending on whether or
not the outer layers are convective, although the $\omega^2$ dividing
$N^2$ makes its effect relatively small), rise rapidly, as though they
are approaching a singularity located at what I call the seismic
surface of the star. This is a structural property of all stars.  In
practice, where the outer layers become optically thin, the
stratification becomes approximately isothermal, and therefore essentially 
exponential, and
the singularity that would be encountered in a polytrope, for example, is
avoided. Nevertheless, for many modes, with $\omega$ rather less than
the value of $\omega_{\rm c}$ in the photosphere, the transition to
exponential behaviour occurs sufficiently well beyond the upper turning point
for its presence to be hardly discernible in the dynamically active
propagating regions below. Therefore the solution to equation (\ref{1.1}) 
actually feels the 
influence of the phantom singularity, and the eigenvalue equation
(\ref{5.5}) needs adjustment. How that is achieved is beyond the scope
of this elementary introduction, and is postponed to a subsequent
article. Here I simply assume that $\omega$ is not so high as to make
that adjustment necessary. There is also a similar problem near the
centre, where $L^2/r^2$ suffers a true (co-ordinate) singularity. This is a
geometrical property of any wave problem in a sphere, and was first
encountered in the context of JWKB theory in quantum mechanics (with a
Coulomb potential). That singularity is well avoided if the degree $l$
of the spherical harmonic (the angular-momentum quantum number in
quantum mechanics) is sufficiently large, so that the lower turning
point $r_1$, which is determined approximately for high-frequency
(ac\-oustic) modes by $r_1/c(r_1)=L/\omega$, is not very close to the
centre. Similar problems face gravity modes, which have very low
frequency, although the problem near the surface of the star is
present only in stars with radiative envelopes, for otherwise the
modes are shielded from the surface by the evanescent convection zone
in which $N^2\le 0$.

With these caveats in mind, I now intrepidly make the substitution
$k\kappa=K$ in equation (\ref{5.5}), using equation (\ref{3.7}) to
obtain
\begin{equation} \label{5.6}
\int^{\tau_2}_{r_1}\left[\frac{\omega^2-\omega^2_{\rm c}}{c^2}
-\frac{L^2}{r^2}
\!\left(\!1-\frac{N^2}{\omega^2}\right)\!\right]^{\frac{1}{2}}\!{\rm d}r
\!\sim\!\left(n-{\textstyle\frac{1}{2}}\right) \pi~,
\end{equation}
hoping that a range of $\omega$ exists which is high enough for the LG
expansion to be reasonably accurate and low enough for the JWKB
approximation near the turning points not to be unduly disturbed by
singularities, phantom or real.

For acoustic modes $\omega^2$ is large, and for many purposes
$N^2/\omega^2$ may be neglected compared with unity, yielding
\begin{equation} \label{5.7}
\frac{\left(n-\frac{1}{2}\right)\pi}{\omega}\sim \int^{\tau_2}_{\tau_1}
\left(1-\frac{\omega^2_{\rm c}}{\omega^2} -\frac{c^2}{r^2
w^2}\right)^{\frac{1}{2}} {\rm d}\tau~,
\end{equation}
where $\tau_1=\tau(r_1),~\tau_2=\tau(r_2)$ and $w=\omega/L$; for
gravity modes $\omega^2/c^2$ may often be neglected, yielding
\begin{equation}
\frac{\left(n-\frac{1}{2}\right)\pi}{L} \sim \int^{r_2}_{r_1}
\left(\frac{N^2}{\omega^2} -1 - \frac{r^2\omega^2_{\rm
c}}{L^2c^2}\right)^{\frac{1}{2}} \frac{{\rm d}r}{r}~,
\end{equation}
in both cases the turning points $r_1$ and $r_2$ being interpreted as
the points at  
which the integrands vanish. 

It is interesting to
develop equation (\ref{5.7}) further in the case when $L$ is large,
for then the lower turning point is close enough to the surface for the
effect of the  
spherical geometry to be small. In that case one may
regard the horizontal phase velocity $v_{\phi\rm h}=rw$ to be essentially
independent of $r$. If, in addition, one approximates the surface
layers of a star of seismic 
radius $R$ by a plane-parallel polytrope of index $\mu$, so
that $c^2=\gamma_1 gz /(\mu+1) $ and $\omega^2_{\rm
c}=\mu(\mu+2)\gamma_1g/4(\mu+1)z$, where $z=R-r$ is depth beneath the
seismic surface of the star (e.g. Gough, 1993), then the acoustic
depth $\tilde{\tau}$ beneath the seismic surface is given by
$\tilde{\tau}=2z/c$, and $\omega^2_{\rm c}=\mu(\mu+2)/\tilde{\tau}^2$.
Equation (\ref{5.7}) reduces to
\begin{eqnarray} \label{67}
\frac{\left(n-\frac{1}{2}\right)\pi}{\omega} &\sim&
\int^{\tilde{\tau}_2}_{\tilde{\tau}_1}
\left[1-\frac{\mu(\mu+2)}{\omega^2\tilde{\tau}^2}
-\frac{\gamma_1^2g^2\tilde{\tau}^2}{4(\mu+1)^2v_{\phi\rm
h}^2}\right]^{\frac{1}{2}} {\rm d}\tilde{\tau} \nonumber \\
&=&\left[\frac{(\mu+1)v_{\phi\rm h}}{2\gamma_1g}
-\frac{{\sqrt{\mu(\mu+2)}}-2}{2\omega}\right]\pi~,
\end{eqnarray}
in which $\tilde{\tau}_1$ and $\tilde{\tau}_2$ are respectively the
acoustic depths of the upper and lower turning points. This equation
may be rewritten
\begin{equation} \label{68}
\frac{(n+\alpha)\pi}{\omega}=F(v_{\phi\rm h})~,
\end{equation}
where $F(v):=(\mu+1)\pi v/2\gamma_1g$ and
$\alpha=\frac{1}{2}[\sqrt{\mu(\mu+2)}-1]=$ constant, which is a special
case of what is now known as Duvall's law (1982), namely equation
(\ref{68}) with the functional form of $F$ unspecified and with the
constant $\alpha$ not necessarily related directly to a polytropic index, and which
formally describes high-frequency acoustic modes in a stellar envelope
with a (well behaved) reflecting surface, whatever its stratification. 
The polytropic form derived here can be
rewritten $\omega^2 \sim(n+\alpha)\tilde{\omega}^2_0 Rk$ as
$k\rightarrow\infty$, where $\tilde{\omega}^2_0 =2\gamma_1 g/(\mu+1)R$
is the square of a characteristic acoustic frequency of the outer
layers of the star, 
and here $k=L/R$ is the horizontal wavenumber at the surface. This is
the parabolic form which approximates the solar $k-\omega$ relation
familiar to helioseismologists, and was used to provide the first
seismic calibration of the adiabatic constant deep in the solar
convection zone. The more general form of equation (\ref{68}) when the
polytropic sound-speed variation is not assumed can be inverted to
give $c/r$ as a function of the observables $(n+\alpha)\pi/\omega$ and
$w$ (e.g. Gough, 1993), and was used to provide the first inferences
of the sound speed through the Sun.

\section{Closing words}

The JWKB  approximation provides  a wonderfully   robust 
representation of
the waves encountered  throughout physics. Turning points are very
common, and, provided one is well clear of them, the approximation
reduces to simple formulae in terms of an exponential function on one
side, and a trigonometrical function on the other, together with the
connexion formula providing the value of the phase of the
trigono\-met\-ri\-cal function according to whether the exponential
function grows or decays away from the turning point. Even near the
turning point, it is straightforward to evaluate the Airy-function
representation from which these simpler formulae are derived.

In solar physics, the leading term in the LG expansion, now commonly
called the WKB approximation, is almost ubiquitous in studies of waves
in the atmosphere, and when turning points are encountered the more
powerful JWKB approximation is available. \,For \,internal
\,acoustic-gravity waves at least one turning point, $r_1$, is always
present, and the connexion formula provided by the JWKB approximation
is essential. I have discussed the latter explicitly, showing how, if
$\omega$ is low enough for the waves to be trapped inside the Sun by
reflection near the surface, the approximate equation (\ref{5.6})
determines the eigenfrequencies of the resonant modes. In the case of
acoustic modes of the Sun and Sun-like stars, that formula is only
approximate. There are two reasons: (i) if $\omega$ is low enough for the
upper turning point not to be too close to the phantom singularity at
the seismic surface of the Sun, then one runs the risk of not having
$k$ large enough for the Airy equation (\ref{4.10}) with the
right-hand side ignored to provide a reliable comparison to the full 
equation. That risk is usually small, because in most (but not all)
cases the Airy-function representation is amazingly robust; (ii) if
the frequency is high, not only does the upper turning point approach
the acoustic surface of the Sun, but it also encounters the upper
superadiabatic boundary layer of the convection zone where, over a
small distance, the acoustic cutoff frequency becomes imaginary; then
there appear to be three turning points which are too close together
for the technique described in \S6 for piecing together different Airy
functions to be applicable. In that case special attention beyond the
scope of this introduction is required, but I might at least point out
that it is pertinent to the interpretation of observational
investigations of the transition to supercriticality, namely to the
high-frequency domain exceeding the acoustic cutoff frequency in the
atmosphere. In the case of gravity waves in a star with a convective
envelope, the threat of a singularity occurs only near the lower
turning point, although in the absence of a convective envelope, the
surface threatens too.

It is worth mentioning that the leading-order LG expansion is formally
applicable only in the limit of infinitesimally slow variation of the
function $f$ in equation (\ref{1.1}). It therefore does not capture
reflection. The JWKB approximation behaves likewise, except near 
the turning point. Reflection in a
spatially varying propagating region is likely to occur everywhere to
some degree, but it is significant only when the scale of variation
$H_f$ of the background state is comparable with or much smaller than
the characteristic inverse wavenumber $(kf^{\frac{1}{2}})^{-1}$ of the
wave. When it is much smaller the variation may be regarded as a
discontinuity, across which one can match two JWKB solutions, in the 
manner adopted by Poisson (1817) under
conditions when the simpler approximation obtained by taking $f$ to be
piecewise constant is \ applicable. \ \  If $kf^{\frac{1}{2}}H_f\approx 1$
over an extended region, one might need to  
adopt a more complicated comparison equation for the
JWKB approximation: to deal with a simple smooth barrier between $z_1$
and $z_2$, for example, one can adopt Weber's equation in the form
$y^{\prime\prime}+k^2(z-z_1)(z-z_2)y=0$ (e.g. Langer, 1959); more
complicated variations in $f$ require correspondingly more complicated
comparison equations. One \,must consider in such cases whether the
result would justify the effort, because direct numerical solution of
an ordinary differential equation is much more straightforward, even
when there are singularities present.

Whether the result justifies the effort depends on the use to which one 
wishes to put the result. If all one needs are eigenfunctions and their
corresponding eigenvalues, the direct numerical computation is in many
cases simpler and more reliable. But analytical results are, for many a 
person, easier to interpret. And then it is more likely that one could 
be able to put 
them to good use. For example, in the early days of helioseismology, the
significance of the so-called large and small frequency separations of
low-degree acoustic modes was first recognized from an asymptotic
analysis, as subsequently, in a sense, was the Duvall law (1982), use of
which has been central to many helioseismological investigations. It
has been argued by some helioseismologists that asymptotic analysis
was actually unnecessary because the discoveries could equally have
been made by numerical investigation. Perhaps they could. But the
fact of the matter is that they were not, and that progress in the
subject would undoubtedly have been slower without asymptotics. That
mode of discovery is likely to continue: the days of asymptotic
analysis in stellar physics are certainly not gone.

\section*{Acknowledgments}
I thank J. B. Keller for useful discussion, and
J. Christensen-Dalsgaard for the hospitality of the Department of
Physics and Astronomy, University of Aarhus , where this
article was written.  Support from the European Helio- and 
Asteroseismology Network (HELAS) for participation at the HELAS 
workshop in Nice is gratefully acknowledged.  HELAS is funded by
the European Union's Sixth Framework Programme.

\end{document}